# Two Paradoxes of the Existence of electric Charge

Claus W. Turtur, University of Applied Sciences Braunschweig-Wolfenbüttel

## Abstract:

Electromagnetic waves propagate with the speed of light. The reason is that electrostatic fields as well as magnetic fields propagate with this speed. Both types of objects, waves as well as static fields contain and transport energy. Consequently it is possible to calculate how much energy and how much energy density a source of a field emits into the space – and the calculation shows that this energy is not zero, for elementary particles as well as for macroscopic spheres. The calculation is presented in this article.
This leads to a principle problem not being answered up to now: From where does the static charge obtain the energy, which it emits permanently ? But the paradox has a second aspect: If we follow the trace of a specified element of volume containing an electric field on its way through the space, we will notice that its contents of field energy decreases during time. But where does this effluent energy go ?

## Structure of the paper:

The article is arranged in two parts: In the first part, the central ideas are presented as an overview but without explanation of details. In the second part, some of the thoughts are explained in detail. The connection between the central ideas in the first part and the associated details in the second part are marked with numbers.

## 1. Part: Overview of the whole contents

<u>1.statement:</u> Electrostatic and magnetostatic fields propagate with the speed of light.
They have the same speed of propagation as electromagnetic waves. Please see detail no.1 as well as the following explanations:
Also static fields can not carry forward their interaction instantaneously – they are also restricted to the speed of light. Of course this upper limit for the speed is clear by principle. But we also know, that the speed of propagation of the static fields really reaches the speed of light (and not a lower speed). This can be understood with the help of the example of the Hertzian dipole emitter, to which further explanations exist in detail no.1.

<u>2.statement:</u> The procedure how an electric charge emits a static field can be compared with the emission of gravitational field from a star in the universe. Both fill the space successively beginning with the moment of their birth – and both do it with the speed of light.

On the basis of the first statement, that electrostatic fields propagate with the speed of light, we perform a thought experiment: We consider a static charge, which has been switched on at the time $t=t_0$, and from that moment on, it emits an electrostatic field. This could be a sphere with a conducting surface, onto which the charge has been brought at the moment $t=t_0$. But the field source can also be an elementary particle, for which the field has been switched "on" in the moment of its genesis, which can be for instance the moment of its production in a laboratory experiment, or for instance it may be the formation of matter after the big bang at the beginning of the universe.

<u>3. statement:</u> Question: From where does the energy emitted with the field originate ?
Let us now turn our attention to a charge with spherical symmetry some time after the field had been switched "on". It permanently emits electrostatic field which contains energy. The problem is: From where does it get this energy ? It does not contain a source of energy and it obviously does not convert mass into energy. Elementary particles (for example electrons) do not alter their mass during time continuously. This is the **first paradox of the existence of electric charge** (as sources of electrostatic fields). A way out of this paradox might be discussed with the assumption, that every electric charge is being permanently supplied with energy from somewhere, but there is no serious indication for such an assumption in physics.

The amount of energy, which an electrically charged particle emits per time interval is calculated in detail no.2. But this calculation leads to a **second paradox of the existence of electric charge:** If we look again at an electric charge with radial symmetry (for instance a charged sphere or a point charge), which emits the electrostatic field with spherical isotropy, we can select a concentric spherical shell of finite thickness and calculate the energy of the electrostatic field within this shell. During time this shell increases its radius, and we can trace it when propagating. Because of the conservation of energy, we expect, that the field energy within this object remains constant during time. But the calculation shows, that this is not the case. The calculation shows, that the empty space reduces the energy within the given spherical shell. This means that the space extracts energy out of the spherical shell. But where does this energy go ? The calculation to which this statement refers, is part of detail no.2.

Now we can combine both paradoxes: On the one hand the charge itself as the source of the field should be supplied permanently with energy, on the other hand the emanating field should give permanently energy to the space. But both energies should not compensate each other completely, because the electrostatic field still contains some energy (as we can even find in textbooks [Fey 01]). By the way the definition of the classical electron's radius goes back to the energy of its electrostatic field.

4. statement: Analogy with gravitation
It should be noticed, that both paradoxes can be found in the static field of gravitation, at least in Newton's conception. Those fields also have a static field energy with its energy density, and they also propagate into the space with the speed of light. This is known from the work on the Detection of Gravitational Waves ([Abr 92], [Ace 02], [And 02], [Bar 99], [Wil 02]). From this point of view, the paradox of the existence of electrostatic charge is completely analogous to the **paradox of the existence of gravitational mass**. Further explanations are to be found in detail no.3.

## 2. part: Explanation of Details

### 1. The speed of propagation of electrostatic fields

It is obvious, that sources of electromagnetic fields and electromagnetic waves permanently emit energy, because the fields as well as the waves contain energy. This argument would already be sufficient to verify the existence of the first paradox, but we want to look at it a bit more precisely, because a more detailed view will bring us later to the second paradox. The consideration begins with a thought experiment in which we can switch "on" and "off" electrostatic fields and magnetic fields. The mechanism for this switching operation can easily be imagined for the example of a magnetic field, because it is easy to switch "on" and "off" an electric current. A possible mechanism to switch "on" and "off" an electrostatic field could be the process of separating differently charged particles from each other inside a closed metal shielding box and taking part of the charge to the outside of the box. But a mechanism, for which a real example exists, is an oscillating pair of electric charges, which acts as a Hertzian dipole emitter. Its mechanism can be understood as following: In the moment in which two electric charges of the same absolute value but of opposite algebraic signs are located at the same position, their electrostatic fields compensate each other exactly. In this moment we regard the electric charge to be switched off, because from every place in the space no charge is observable. As soon as both charges get different positions, as they do when they oscillate with regard to each other, each of them generates an electric field, and as long as both have different positions, they produce different field strengths at every point in the space. The superposition of these both fields is different from zero. During this time of the oscillation, the charge can be regarded to be switched "on". Only during those moments when their positions coincide, the electric field is switched "off". In a similar way, both charges generate a magnetic field as long as they are moving, but in the reversal point of the oscillation, when the movement stops for a moment, no magnetic field is emitted.

And all these fields, which are already generated and emitted, propagate into the space with the speed of light, without paying any attention to the question, whether there is some other field emitted earlier or later or not. As long as the oscillation of the both charges is going on, electric and magnetic fields with alternating field strength is emitted and propagate into the space. This is the typical explain of the Hertzian dipole emitter with its characteristics.

Normally on the basis of these considerations, the emission characteristics of the Hertzian dipole emitter follows from the knowledge of the positions and the speed of the electrical charges as a function of time during their oscillation by calculating the electrostatic and the magnetic field strength at every time and every point in the space (according to Coulomb's law and according to Biot-Savart's law) and taking the speed of propagation of these fields into account, which is the speed of light.

In addition to the calculation, there is also a plausible illustration for the emission characteristics of the Hertzian dipole emitter:

In that part of the oscillation in which the field sources are on a movement from the common center away, the field sources follow the emitted field (which is also going from the common center away) and by that they enhance the electric field strength. In the opposite part of the period of the oscillation, in which the field sources are on their movement towards the common center, they go away from the emitted field (resp. they move towards the field into the opposite direction) and by that they supply less field strength per volume to the outside of the oscillation (and the field into the opposite direction is less strong because of the distance from the common center).

The magnetic field is maximal in those moments in which the field sources pass the common center, because in those moments the speed of the field sources reaches its maximum. In the contrary to that, the moments of maximal deflection correspond to the minimal speed of movement of the field sources (in these moments their speed is zero), so that at this time no magnetic field is generated. By the way the phasing between the electric and the magnetic field occurring at electromagnetic waves is also explained from this consideration.

## 2. Calculation of the emitted energy and the propagation of this energy into the space

Electric charges with spherical symmetry emit the field with spherical symmetry, and so it should be imaginable to calculate the field energy which the surface emits per time interval. For charges spheres with finite size, this way of calculation would be fine, but for punctiform charges it is not possible to define their surface or radius. One of them is the electron. Its classical radius ($r_{klass.} = 2.8179... \cdot 10^{-15} m$ [Cod 00]) is defined by the means of its field energy but the value is in contradiction to the observations of scattering experiments, which lead to an upper limit of the electron's radius below $10^{-18} m$ (see e.g. [Loh 05]). Because it is not helpful to deal with this problem here, we surround the field source "Q" with a spherical shell and calculate the energy passing the shell. The inner radius of the shell is not important, because the energy coming from the field source has to pass the shell. This is the case for elementary particles in the same way as for macroscopic spheres. For the illustration of this statement please look to fig.1. The interpretation is the following:

- We start our considerations at the time $t=0$ at which the field has just filled the inner shell with the radius $x_1$. Let us compare this with a moment $\Delta t > 0$ following a bit later. At this moment the field reaches the shell with the radius $x_1 + c \cdot \Delta t$ ($c$ = speed of light), so that during the time interval $\Delta t$ the emitted energy has the same amount as the energy within the shell from $x_1$ to $x_1 + c \cdot \Delta t$, because the total energy of the total field was enhanced just by this amount. This amount of energy, which is not zero can only by generated by the source in the center of the sphere, because there is no other source. This clarifies that the question about the origin of this energy is valid and leads to the first paradox. Subsequent to fig.1 the emitted energy will be calculated.

- In a second step we want to consider an even more later moment of time $t_2$ at which the field fills a sphere of the radius $x_2$. And we want to calculate the energy which was emitted by the source within the time interval from $t_2$ to $t_2 + \Delta t$. This is the energy within the shell from $x_2$ to $x_2 + c \cdot \Delta t$. As long as the empty space is not a source or a drain of energy and the source emits constantly with time, the amount of energy within the outer shell should be the same as the amount of energy within the inner shell. The identity of the amount of energy within both spherical shells can also be understood because of the fact, that the shell has moved within the time interval $t_2$ to the position of the outer shell, and it should no alter the amount of energy it contains just by simple moving to a new position. But if the energy in both shells (the inner and the outer shell) differ from each other (this will be confirmed by the calculation following soon), it is clear that we have to ask where this energy is going to. This is the reason for the paradox no.2.

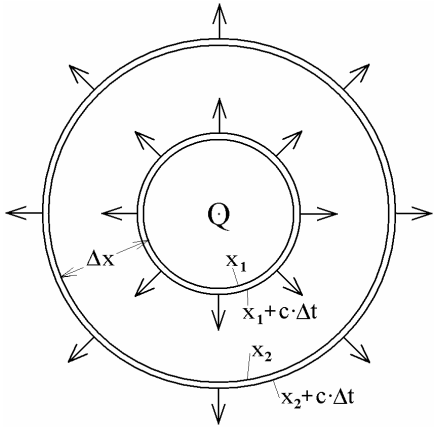

**Fig.1:**
Illustration of a spherical shell which contains a certain amount of field energy. The sense of this construction is to trace the field energy when passing the empty space.

The calculation of the amount of energy, which must be done to understand the paradoxes, is following now:

- The field strength produced by a charge $Q$ with radial symmetry (i.e. a punctiform charge or a charge with spherical symmetry) is according to Coulomb's law $\vec{E}(\vec{r}) = \frac{1}{4\pi\varepsilon_0} \cdot \frac{Q}{r^3} \cdot \vec{r}$, where the center of the charge is located in the origin of coordinates and $\vec{r}$ is the position of an arbitrary point in the space at which the field strength shall be determined.

- If we write $\vec{r}$ in spherical coordinates with $\vec{r} = (r, \vartheta, \varphi)$, the absolute values of the field strength are dependant not of the direction of $\vec{r}$ but only of the absolute value of $r = |\vec{r}|$, namely $E = |\vec{E}| = \frac{1}{4\pi\varepsilon_0} \cdot \frac{Q}{r^2}$.

- The energy density of the electric field is $u = \frac{\varepsilon_0}{2} \cdot |\vec{E}|^2$.

- Consequently the energy density of the field produced by a charge with spherical symmetry is $u = \frac{\varepsilon_0}{2} \cdot |\vec{E}|^2 = \frac{\varepsilon_0}{2} \cdot \left( \frac{1}{4\pi\varepsilon_0} \cdot \frac{Q}{r^2} \right)^2 = \frac{Q^2}{32\pi^2\varepsilon_0 r^4}$.

- The energy within the spherical shell from $x_1$ to $x_1 + c \cdot \Delta t$ can now be calculated as the Volume integral

$$E_{inner \atop shell} = \int_{spherical \atop shell} u(\vec{r}) dV = \int_{\varphi=0}^{2\pi} \int_{\vartheta=0}^{\pi} \int_{r=x_1}^{x_1+c\cdot\Delta t} \frac{Q}{32\pi^2\varepsilon_0 r^4} \cdot r^2 \cdot \sin(\vartheta)\, dr\, d\vartheta\, d\varphi$$

$$= \frac{Q}{32\pi^2\varepsilon_0} \cdot \int_{\varphi=0}^{2\pi} \int_{\vartheta=0}^{\pi} \underbrace{\int_{r=x_1}^{x_1+c\cdot\Delta t} \frac{1}{r^2} \cdot dr}_{=\frac{c\cdot\Delta t}{(x_1+c\cdot\Delta t)\cdot x_1}} \cdot \sin(\vartheta)\, d\vartheta\, d\varphi$$

$$= \frac{Q}{32\pi^2\varepsilon_0} \cdot \frac{c\cdot\Delta t}{(x_1+c\cdot\Delta t)\cdot x_1} \cdot \underbrace{\int_{\varphi=0}^{2\pi} \underbrace{\int_{\vartheta=0}^{\pi} \sin(\vartheta)\, d\vartheta}_{=2}\, d\varphi}_{=4\pi}$$

$$= \frac{Q}{32\pi^2\varepsilon_0} \cdot \frac{c\cdot\Delta t}{(x_1+c\cdot\Delta t)\cdot x_1} \cdot 4\pi = \frac{Q}{8\pi\varepsilon_0} \cdot \frac{c\cdot\Delta t}{(x_1+c\cdot\Delta t)\cdot x_1}$$

Obviously this energy is not zero. This means that the source indeed emits energy permanently. This is the reason for the first paradox.

- Let now the time elapse until it reaches $t_2$. The inner border of the observed spherical shell has now passed from $x_1$ to $x_2$ and the outer border from $x_1 + c\cdot\Delta t$ to $x_2 + c\cdot\Delta t$. With the distance $\Delta x$ introduced in fig.1, we find the inner and the outer border of the shell being at the radii $x_2 = x_1 + \Delta x$ respectively $x_2 + c\cdot\Delta t = x_1 + \Delta x + c\cdot\Delta t$. The spherical shell has enhanced its volume, but the field strength within this moving shell has been reduced (in accordance with Coulomb's law). If the empty space would allow the field energy just to pass by, the amount of energy within the outer shell $E_{outer\ shell}$ should be the same as the amount of energy within the inner shell $E_{inner\ shell}$. We want to check this:

$$E_{outer \atop shell} = \int_{sherical \atop shell} u(\vec{r}) dV = \int_{\varphi=0}^{2\pi} \int_{\vartheta=0}^{\pi} \int_{r=x_2}^{x_2+c\cdot\Delta t} \frac{Q}{32\pi^2\varepsilon_0 r^4} \cdot r^2 \cdot \sin(\vartheta)\, dr\, d\vartheta\, d\varphi$$

$$= \frac{Q}{32\pi^2\varepsilon_0} \cdot \int_{\varphi=0}^{2\pi} \int_{\vartheta=0}^{\pi} \underbrace{\int_{r=x_1+\Delta x}^{x_1+\Delta x+c\cdot\Delta t} \frac{1}{r^2} \cdot dr}_{=\frac{c\cdot\Delta t}{(x_1+\Delta x+c\cdot\Delta t)\cdot(x_1+\Delta x)}} \cdot \sin(\vartheta)\, d\vartheta\, d\varphi$$

$$= \frac{Q}{32\pi^2\varepsilon_0} \cdot \frac{c\cdot\Delta t}{(x_1+\Delta x+c\cdot\Delta t)\cdot(x_1+\Delta x)} \cdot \underbrace{\int_{\varphi=0}^{2\pi} \underbrace{\int_{\vartheta=0}^{\pi} \sin(\vartheta)\, d\vartheta}_{=2} d\varphi}_{=4\pi}$$

$$= \frac{Q}{32\pi^2\varepsilon_0} \cdot \frac{c\cdot\Delta t}{(x_1+\Delta x+c\cdot\Delta t)\cdot(x_1+\Delta x)} \cdot 4\pi = \frac{Q}{8\pi\varepsilon_0} \cdot \frac{c\cdot\Delta t}{(x_1+\Delta x+c\cdot\Delta t)\cdot(x_1+\Delta x)}$$

- Obviously the energy $E_{inner\ shell}$ is more than the energy $E_{outer\ shell}$. This means that the empty space decreases the energy of the shell. This proofs the validity of the second paradox, which puts the question: Where does the difference of energy go ? In order to complete the thoughts, fig.2 shows a plot of the difference of energy $E_{diff} = E_{inner\ shell} - E_{outer\ shell}$, which the spherical shell lost on its way $\Delta x$ across the empty space, beginning from the radius $x_1$.

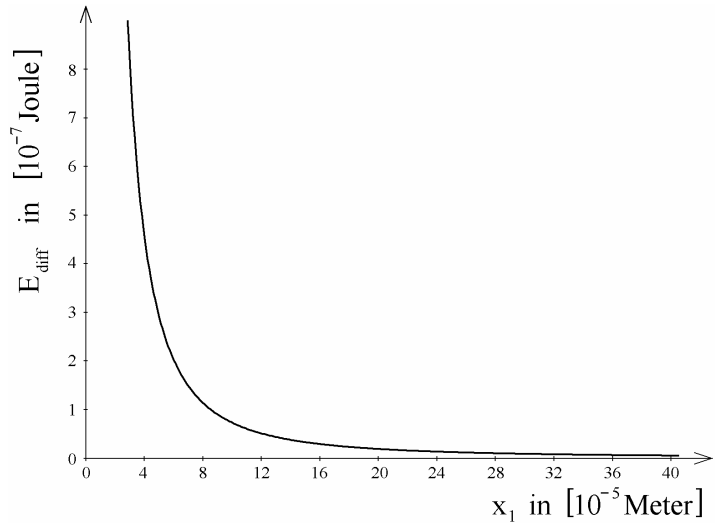

**Fig.2:**
Plot of the energy difference which the spherical shell of our thought experiment lost when passing the empty space.
$x_1$ is the inner radius of the shell.
For the example of our calculation the input was taken as $\Delta x = 10^{-10} m$ and $\Delta t = 10^{-7}$ sec., and the source of the field was an electron.

## 3. The analogy of the problem for the case of gravitation

At least in Newton's description of gravitation we see the following conception:
The analogy of the **paradox of the existence of electric charge** with the **paradox of the existence of gravitational mass** can be understood by the means of the analogy of the propagation of static fields and waves.
The gravitational counterpart of the electrostatic field is the field of gravitation. The gravitational counterpart of electromagnetic waves are gravitational waves.[1] Their existence is accepted today, and the work is in process to detect them experimentally.
Also the mechanism of their generation can be understood in analogy with the generation of electromagnetic waves. In the electromagnetic case, oscillating charges can be responsible for periodic switching "on" and "off" the field an by that they generate the waves. In the case of gravitation we know for instance the example of a double star (or other examples) [Sha 83] rotating around a common center of mass, and creating gravitational waves. The characteristics of the emitted field looks different from the characteristics created by a harmonic oscillation, but this does not influence the principle aspects of our logical track here. The important aspect is that it is generally accepted, that gravitational waves together with gravitational fields propagate with the speed of light.
Thus the analogy with the mechanism and the formula of detail 2 is obvious. We just have to replace the energy density of the electrostatic field by the energy density of the gravitational field, which is $u = \frac{1}{8\pi\gamma} \cdot |\vec{G}|^2$ (with $\gamma$ = Newton's constant of gravitations and $\vec{G} = \gamma \cdot \frac{M}{r^3} \cdot \vec{r}$, where $M$ is the gravitational mass) and put this energy density into volume integrals shown there. For the calculation, only the values of the constants in front of the integrals have to be changed in appropriate manner, so that the calculation does not have to be demonstrated once more now. The crucial point is clear:
Also for gravitational fields the two paradoxes remain as open questions:
- From where does the energy descend which every gravitational mass emits permanently ?
- Why does the empty space absorb energy from the gravitational field during propagation – and where does this energy go ?

---

[1] By the way it could be mentioned that there is a gravitational counterpart to the magnetic field, which is called gravimagnetic field. Its background is the Thirring-Lense-effect (see e.g. [Sch 02], [Thi 18], [Gpb 07]). But this gravimagnetic field is not necessary to understand the paradoxes explained here, and so we want to omit it now.

**Literature**

[Abr 92]   "The Laser interferometer gravitational wave observatory"
A. Abramovici et. al., Science 256, S.325-333 (1992)

[Ace 02]   "Status of the VIRGO"
F. Acernese et. el., Classical and Quantum Gravity 19, S.1421 (2002)

[And 02]   "Current status of TAMA", M. Ando and the TAMA collaboration,
Classical and Quantum Gravity 19, S.1409 (2002)

[Bar 99]   "LIGO and the Detection of Gravitational Waves"
R.Barish and R.Weiss, Phys. Today 52 (Oct), 44 (1999)

[Cod 00]   "CODATA Recommended Values of the Fundamental Physical Constants: 1998"
Review of Modern Physics 72 (2) 351 (April 2000).
The contents of CODATA is permanently updated at:
http://physics.nist.gov/cuu/Constants/

[Fey 01]   "Feynman Vorlesungen über Physik, Band II: Elektromagnetismus und
Struktur der Materie", R. P. Feynman, R. B. Leighton und M.Sands, (2001),
3.Auflage, Oldenbourg Verlag, ISBN 3-486-25589-4

[Gpb 07]   "Gravity- Probe- B  Experiment", Stanford- University, F. Everitt et. al.
to be found in (2007) at:   http://einstein.stanford.edu/index.html

[Loh 05]   "Hochenergiephysik", Erich Lohrmann, (2005)
B. G. Teubner Verlag, ISBN 3-519-43043-6

[Sch 02]   "Gravitation", U. E. Schröder (2002)
Wissenschaftlicher Verlag Harri Deutsch GmbH, ISBN 3-8171-1679-9

[Sha 83]   "Black Holes, White Dwarfs and Neutron Stars: The Physics of Compact Objects"
Stuart L. Shapiro und Saul A. Teukolsky, (1983)
Wiley Interscience ISBN 0-471-87317-0

[Thi 18]   "Über die Wirkung rotierender ferner Massen in Einsteins Gravitationstheorie"
von Thirring und Lense, Phys. Zeitschr. 19, Seiten 33-39 Jahrgang (1918)

[Wil 02]   "A report of the status of the GEO 600 gravitational wave detector"
von B.Willke et. al., Classical and Quantum Gravity. 19, S.1377 (2002)

**Author's Adress:**
Prof. Dr. Claus W. Turtur
University of Applied Sciences Braunschweig-Wolfenbüttel
Salzdahlumer Straße 46 / 48
Germany - 38302 Wolfenbüttel
Email: c-w.turtur@fh-wolfenbuettel.de
Tel.: (++49) 5331 / 939 – 3412